\documentstyle[epsfig]{mn}

\newif\ifAMStwofonts

\ifoldfss
  \ifCUPmtlplainloaded \else
    \NewTextAlphabet{textbfit} {cmbxti10} {}
    \NewTextAlphabet{textbfss} {cmssbx10} {}
    \NewMathAlphabet{mathbfit} {cmbxti10} {} 
    \NewMathAlphabet{mathbfss} {cmssbx10} {} 
  \fi
  \ifAMStwofonts
    \ifCUPmtlplainloaded \else
      \NewSymbolFont{upmath} {eurm10}
      \NewSymbolFont{AMSa} {msam10}
      \NewMathSymbol{\upi}     {0}{upmath}{19}
      \NewMathSymbol{\umu}     {0}{upmath}{16}
      \NewMathSymbol{\upartial}{0}{upmath}{40}
      \NewMathSymbol{\leqslant}{3}{AMSa}{36}
      \NewMathSymbol{\geqslant}{3}{AMSa}{3E}

    \fi
  \fi
\fi 

\ifnfssone
  \newmathalphabet{\mathit}
  \addtoversion{normal}{\mathit}{cmr}{m}{it}
  \addtoversion{bold}{\mathit}{cmr}{bx}{it}
  \newmathalphabet{\mathbfit} 
  \addtoversion{normal}{\mathbfit}{cmr}{bx}{it}
  \addtoversion{bold}{\mathbfit}{cmr}{bx}{it}
  \newmathalphabet{\mathbfss} 
  \addtoversion{normal}{\mathbfss}{cmss}{bx}{n}
  \addtoversion{bold}{\mathbfss}{cmss}{bx}{n}
  \ifAMStwofonts
    \ifCUPmtlplainloaded \else
      \UseAMStwoboldmath
      \makeatletter
      \new@mathgroup\upmath@group
      \define@mathgroup\mv@normal\upmath@group{eur}{m}{n}
      \define@mathgroup\mv@bold\upmath@group{eur}{b}{n}
      \edef\UPM{\hexnumber\upmath@group}
      \new@mathgroup\amsa@group
      \define@mathgroup\mv@normal\amsa@group{msa}{m}{n}
      \define@mathgroup\mv@bold\amsa@group{msa}{m}{n}
      \edef\AMSa{\hexnumber\amsa@group}
      \makeatother
      \mathchardef\upi="0\UPM19
      \mathchardef\umu="0\UPM16
      \mathchardef\upartial="0\UPM40
      \mathchardef\leqslant="3\AMSa36
      \mathchardef\geqslant="3\AMSa3E
    \fi
  \fi
\fi 

\ifnfsstwo
  \DeclareMathAlphabet{\mathbfit}{OT1}{cmr}{bx}{it}
  \SetMathAlphabet\mathbfit{bold}{OT1}{cmr}{bx}{it}
  \DeclareMathAlphabet{\mathbfss}{OT1}{cmss}{bx}{n}
  \SetMathAlphabet\mathbfss{bold}{OT1}{cmss}{bx}{n}
  \ifAMStwofonts
    \ifCUPmtlplainloaded \else
      \DeclareSymbolFont{UPM}{U}{eur}{m}{n}
      \SetSymbolFont{UPM}{bold}{U}{eur}{b}{n}
      \DeclareSymbolFont{AMSa}{U}{msa}{m}{n}
      \DeclareMathSymbol{\upi}{0}{UPM}{"19}
      \DeclareMathSymbol{\umu}{0}{UPM}{"16}
      \DeclareMathSymbol{\upartial}{0}{UPM}{"40}
      \DeclareMathSymbol{\leqslant}{3}{AMSa}{"36}
      \DeclareMathSymbol{\geqslant}{3}{AMSa}{"3E}
    \fi
  \fi
\fi 

\ifCUPmtlplainloaded \else
  \ifAMStwofonts \else 
    \def\upi{\pi}
    \def\umu{\mu}
    \def\upartial{\partial}
  \fi
\fi

\title{Temperature effects on the 15-85-$\mu$m Spectra of Olivines and Pyroxenes} 
\author[J. E. Bowey et
al.]{J. E. Bowey$^{1,2}$, C. Lee$^{1}$, C. Tucker$^{1}$,
A. M. Hofmeister$^{3}$, P. A. R. Ade$^{1}$ and M. J. Barlow$^2$\\
$^1$Dept. of Physics, Queen Mary, University of London, Mile End Road,
London, E1 4NS, UK.\\ $^2$Dept. of Physics \& Astronomy, University
College London, Gower Street, London, WC1E 6BT (e-mail:
jeb@star.ucl.ac.uk)\\ $^3$Dept. Earth $\&$ Planet. Sci., Washington
University, 1 Brookings Dr., St Louis, MO 63110, USA.}

\date{Accepted ???. Received ???; in original form ???}
\pagerange{\pageref{firstpage}--\pageref{lastpage}}
\pubyear{???}

\begin{document}
\maketitle 
\label{firstpage}
\begin{abstract}
Far-infrared spectra of laboratory silicates are normally obtained at
room temperature even though the grains responsible for astronomical
silicate emission bands seen at wavelengths $>20~\mu$m are likely to
be at temperatures below $\sim 150$~K. In order to investigate the
effect of temperature on silicate spectra, we have obtained absorption
spectra of powdered forsterite and olivine, along with two
orthoenstatites and diopside clinopyroxene, at 3.5$\pm0.5$~K and at
room temperature (295$\pm2$~K). To determine the changes in the
spectra the resolution must be increased from $\sim 1$ to
0.25~cm$^{-1}$ at both temperatures since a reduction in temperature
reduces the phonon density, thereby reducing the width of the infrared
peaks. Several bands observed at 295~K split at 3.5~K. At 3.5~K the
widths of isolated single bands in olivine, enstatites and diopside
are $\sim 90\%$ of their 295~K-widths. However, in forsterite the
3.5-K--widths of the 31-, 49- and 69-$\mu$m bands are, respectively,
90\%, 45\% and 31\% of their 295~K widths. Due to an increase in
phonon energy as the lattice contracts, 3.5-K--singlet peaks occur at
shorter wavelengths than do the corresponding 295-K peaks; the
magnitude of the wavelength shift increases from $\sim 0-0.2~\mu$m at
25~$\mu$m to $\sim 0.9~\mu$m at 80~$\mu$m. In olivines and enstatites
the wavelength shifts can be approximated by polynomials of the form
$ax+bx^2$ where $x=\lambda_{pk}(295$~K$)$ and the coefficients $a$ and
$b$ differ between minerals; for diopside this formula gives a lower
limit to the shift. Changes in the relative absorbances of spectral
peaks are also observed. The temperature dependence of $\lambda_{pk}$
and bandwidth shows promise as a means to deduce characteristic
temperatures of mineralogically distinct grain populations.  In
addition, the observed changes in band strength with temperature will
affect estimates of grain masses and relative mineral abundances
inferred using room-temperature laboratory data.  Spectral
measurements of a variety of minerals at a range of temperatures are
required to fully quantify these effects.
\end{abstract}


\section{INTRODUCTION}
Mid and far-infrared spectra obtained with the Infrared Space
Observatory Short- and Long-Wavelength Spectrometers (SWS and LWS,
respectively) have revealed emission bands which have been associated
with crystalline silicate dust. The environments include comets, young
stellar objects and oxygen-rich dust in outflows and disks associated
with late-type stars and planetary nebulae (e.g. Crovisier
et~al.~2000, Malfait et~al.~1999, Waters et~al.~1996, Sylvester et
al.~1999 and Cohen et~al.~1999, respectively). Simple spectral fits
indicate that much of the optically thin silicate emission occurs at
temperatures in the 50-100~K range (e.g. Sylvester et~al. 1999). However,
previous laboratory studies (e.g. Mennella et~al.~1998, Henning \&
Mutschke~1997, Day~1976, Agladze et~al.~1996) have shown that the
optical properties of silicates vary between 295~K and 10~K. In MgO
the frequency and FWHM($\nu$) of the 24.5~$\mu$m (295K) TO mode
(Jasperse et al. 1964) vary nearly linearly with temperature between
7.5 and 1950~K (Kachare et al. 1972). A similar study of the effect of
423--873K-temperatures on 3-12$\mu$m PAH bands by Joblin et al. (1995)
revealed changes in band structure, bandwidth and frequency which
indicate that free (as opposed to condensed) PAH molecules carry the
astronomical 3.3$\mu$m band.

Due to a lack of laboratory data, both simple black body fits and
radiative transfer models of these environments have used silicate
laboratory spectra obtained at room temperature. In addition to
fitting ISO data, spectra obtained for a range of samples and
temperatures will also be required for comparison with data from
SIRTF, FIRST and the NGST. To begin to quantify the effect of
temperature on silicates we compare spectra of two olivines, two
orthoenstatites and diopside clinopyroxene at 295~K and 3.5~K.

\subsection{Choice of Spectral Resolution}
At cryogenic temperatures the vibrational states of a solid are
substantially depopulated in comparison to their values at 295~K in
accord with statistical thermodynamics.  From the damped harmonic
oscillator model, the width of a peak is related to the lifetime
between phonon scattering events
\cite{jebwooten_1972,jebhofmeister_1999}.  As the number of states
decreases, the possibilities of scattering decrease, and the peaks
must therefore sharpen. The vibrational bands observed in cryogenic
$10-100~\mu$m spectra of \emph{crystalline} silicates
\cite{jebday_1976,jebhenning_1997} are sharper and deeper than those
obtained at room temperature. Medium ($\sim$1-2~cm$^{-1}$) resolution
spectra (Henning \& Mutschke 1997, Mutschke pers. commun. and Mennella
et~al. 1998) indicate that the bands of crystalline silicates shift in
wavelength as the temperature is reduced, and that the magnitude of
the shift increases with wavelength. If the spectral resolution is
insufficient, very narrow peaks are under-sampled, leading to rounded
profiles and incorrect measurements of the wavelength
shifts. Therefore, higher-resolution measurements are crucial in the
far-IR where lower-resolution data (Mennella et al.~1998) showed that
the isolated $\sim 70~\mu$m (143 ~cm$^{-1}$) peak of forsterite has a
24~K width of $\sim$4~cm$^{-1}$ and a $\sim$2~cm$^{-1}$ shift between
295~K and 24~K. We therefore present 0.25~cm$^{-1}$-resolution
transmission spectra of crystalline minerals from the olivine and
pyroxene groups at 3.5~K and 295~K.

\section{Lattice structures and 295~K Band Assignments}
\subsection{Olivines}
\begin{figure} \begin{center}
\epsfig{file=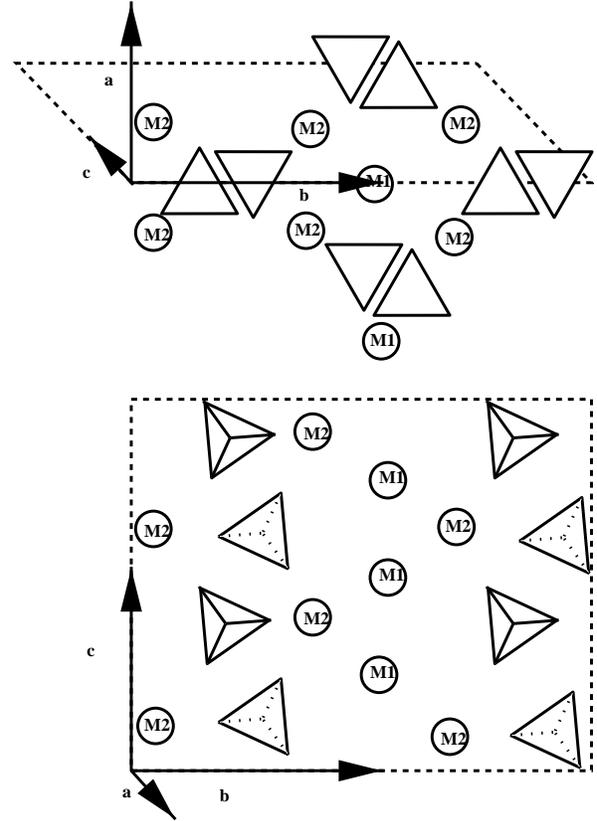,height=11cm,clip=,bbllx=0,bburx=372,bblly=0,bbury=530}
\end{center}
\caption{Schematic three-dimensional structure of crystalline olivines. The lower diagram is a slice through the upper diagram along the \emph{b} axis. Tetrahedra represent SiO$_4^{4-}$ groups whose apices point out of the page (solid) and into the page (dotted). M1 and M2 represent the sites of divalent metal cations.}
\label{fig:olivstruc}
\end{figure}

The general formula for the olivine solid solution series is
(Mg$_x$Fe$_{1-x})_2$SiO$_4$; the Mg$^{2+}$ end-member is forsterite,
the Fe$^{2+}$ end member is fayalite. Olivine minerals have the
structure shown in Figure \ref{fig:olivstruc}. In the figure each
tetrahedron represents the SiO$_4^{4-}$ anion which consists of an Si
atom surrounded by oxygen atoms at each corner of the tetrahedron; M1
and M2 represent the sites of the metal cations. In end-member
forsterite both sites are occupied by Mg$^{2+}$. In olivine some of
the Mg$^{2+}$ in sites M1 and M2 is replaced by Fe$^{2+}$; in this
case there is no preference for substitution between the sites since
these two ions have similar radii. The active modes in the
15-85-$\mu$m region are (after Hofmeister~1997): bending of the
tetrahedral O--Si--O ($\lambda(295$K$)\la 20~\mu$m), rotation of the
SiO$_4^{4-}$ tetrahedra (four bands between 21 and 26~$\mu$m),
translation of the SiO$_4^{4-}$ tetrahedra ($\sim 50~\mu$m) and
translation of the Mg$^{2+}$ and Fe$^{2+}$ cations (several of the
bands in the 21--30-$\mu$m region); two bands are associated with
combined translations of the divalent cations and SiO$_4^{4-}$
tetrahedra (at $\sim 36.5$ and $\sim 70~\mu$m in forsterite; at $\sim
94$ and $\sim 109~\mu$m in fayalite, Fe$_2$SiO$_4$).

\subsection{Ortho- and Clino-pyroxenes}
\begin{figure}
  \begin{center}
    \epsfig{file=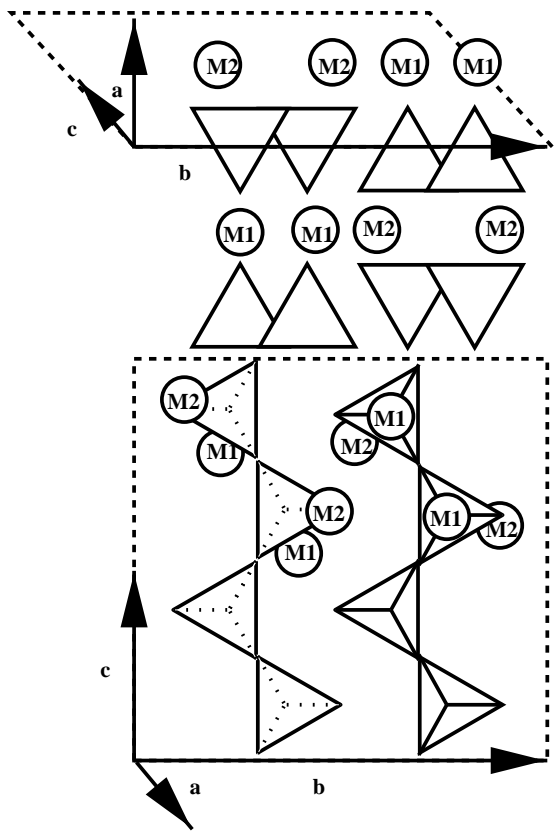,height=11cm,clip=,bbllx=0,bburx=290,bblly=0,bbury=420}
  \end{center}
\caption{Schematic three-dimensional structure of ortho-enstatite. The lower diagram is a slice through the upper diagram along the \emph{b} axis. The $b-c$ plane is characterized by chains of silicate tetrahedra joined by shared oxygen atoms. These are surrounded by metal ions (for clarity these are omitted from the lower pairs of tetrahedra in each chain). In the $a-b$ plane M1 and M2 sites occur, respectively, at the apices and bases of the silicate tetrahedra. M2 sites are larger than M1 sites; therefore in diopside-like clinopyroxenes relatively large Ca$^{2+}$ ions can occupy M2 sites but only smaller
Mg$^{2+}$ and Fe$^{2+}$ ions may occupy the M1 sites.}
\label{fig:pyroxstruc}
\end{figure}

Minerals in the pyroxene solid solution series
(Mg$_x$,Fe$_y$,Ca$_z$)$_2$Si$_2$O$_6$ where $x+y+z=1$, belong to one
of two crystal systems: orthorhombic, in which the three crystal axes
are orthogonal, and monoclinic, which has two orthogonal and one
inclined axis. The structure of orthopyroxene is shown in Figure
\ref{fig:pyroxstruc}; silicate tetrahedra are joined together by
shared oxygen atoms to form long chains along the $c$ axis. Once
again, Mg$^{2+}$ and Fe$^{2+}$ can replace each other in the M1 and M2
sites. The precise band assignments for the orthopyroxenes are
unknown. However, comparison with the bands of the clinopyroxenes and
olivines suggests that the bands longwards of $20~\mu$m may involve
translations of the cations, whereas bands at wavelengths shorter than
$20~\mu$m involve the deformation of the tetrahedral and
inter-tetrahedral O--Si--O bonds.
 
In clinopyroxenes similar to diopside (there are other types of
clinopyroxene), the larger Ca$^{2+}$ ion ($r\sim 1.12$\AA) occurs only
in the M2 sites, whilst the smaller Mg$^{2+}$ ($r\sim 0.72$\AA) and
Fe$^{2+}$ ($r\sim 0.78$\AA) ions occur in both these and the smaller
M1 sites.  This changes the alignment of the tetrahedra, resulting in
a crystal shape with an inclined $a$ axis. Band assignments in
diopsides are currently derived by comparison between force-constant
calculations and laboratory data. The active modes in the 15-85-$\mu$m
region are modelled as (Tomisaka \& Iishi 1980): bending of the
tetrahedral and inter-tetrahedral O--Si--O bonds shortwards of 30~$\mu$m
at 295~K (at 23.2 and 29.6~$\mu$m these may be blended with Mg$^{2+}$
translations); Mg$^{2+}$ translations blended with translations of the
O--Si--O chain and rotations of the tetrahedra at 32.3~$\mu$m; Mg$^{2+}$
translations blended with tetrahedral rotations at 34.3$~\mu$m;
Ca$^{2+}$ translations blended with tetrahedral rotations at
40.4~$\mu$m; Ca$^{2+}$ translations at 44.9~$\mu$m; finally a broad
band at $\sim 66~\mu$m is produced by a zig-zag motion of bridging
oxygen atoms in the silicate chain. However, force constant models are
not backed-up by laboratory measurements, and the assignments are
tentative.  Measurements of the effect of isotopic or chemical
substitutions are needed for reliable band assignments.

\section{THE EFFECT OF TEMPERATURE}
\subsection{Changes in bond length}
\label{sec:bondlength}
If the bond length decreases, the energy of the associated phonon
increases. (This physics is equivalent to the energy levels of a
particle in a box; the effect is frequently observed in pressure
studies, e.g. Ferraro 1984).  Because frequency increases as pressure
increases (or temperature decreases), the wavelength of the infrared
peak must decrease. Tetrahedral Si--O bond lengths and bond angles are
relatively incompressible in comparison to the spacings between the
oxygen and metal ions and inter-tetrahedral oxygen atoms so there is
little variation in the wavelength of the 10-$\mu$m stretching and
20-$\mu$m bending features with temperature (e.g. Day~1976). In
contrast, the volumes of the divalent-cation sites are more
compressible. Therefore one would expect to see a greater change in
the wavelengths of peaks associated with the cation sites.

\subsection{Changes in band occupancy}
\label{sec:bandocc}
The lifetime, $\Gamma$, of a phonon (the time between scattering
events) is given by $\Gamma^{-1} \propto 2 \pi \Delta\nu(\epsilon_2)$,
where $\Delta\nu(\epsilon_2)$ is the full-width-half-maximum of the peak
in the imaginary part of the dielectric function, $\epsilon_2$, in
frequency units. (Wooten~1972). When the bands are narrow and weak,
the frequency for the atoms vibrating parallel to the propagation
direction of the radiation (the longitudinal optic; LO) is similar to
the frequency of the atoms vibrating perpendicular to the propagation
direction (the transverse optic; TO), thereby creating a symmetrical
absorption peak whose shape approaches the Lorentzian shape of its
counterpart in $\epsilon_2$ (Hofmeister and Mao 2001). Absorbance
full-width-half-maxima (FWHM) are related to dielectric widths;
therefore the widths of the absorbance peaks should be roughly
proportional to the dielectric widths in wavenumbers.

At temperature, $T$, the occupation of phonon energy levels relative
to the ground state is given by the partition function
$q=\sum_{n=0}^\infty\exp{[-\frac{n h\nu}{kT}]}$, where $k$ is the
Boltzmann constant and it is assumed that the phonon frequency, $\nu$,
is not a function of temperature (this is not true; see Section
\ref{sec:bondlength}). As the temperature is decreased the number of
vibrational states decreases and the numbers of phonon-phonon
scattering events decrease so that phonon scattering off the crystal
sides and defects becomes more important. For phonon-phonon
scattering, the lifetime is inversely proportional to the number of
phonons available, and hence as the sample is cooled, states are
depopulated, and the lifetime increases and peak width decreases. In
practise, FWHMs are not often measured and the FWHM temperature
dependence is not well understood (see Hofmeister, in prep).

\section{THE EXPERIMENT}
\subsection{Sample Preparation}
\begin{table*}
\caption{Origin and composition of crystalline samples}
\label{tab:samples}
\begin{minipage}{\linewidth}
\begin{tabular}{llll}
Group&Mineral&Composition&Source\\
Olivine&Synthetic Forsterite\footnote{Synthesized from oxide powders by
H. K. Mao}&Mg$_2$SiO$_4$&Geophysical Laboratory,
\\ &&&Washington,
D. C., USA.\\&Olivine (var. peridot)&
(Mg$_{1.77}$Fe$_{0.20}$Al$_{0.01}$Ni$_{0.01}$)SiO$_4$& Navajo Indian
Reservation\\ 
&&&San Carlos, Arizona, USA\\
 &&&\\ 
Orthopyroxene
&Enstatite&(Mg$_{1.8}$Fe$_{0.16}$Ca$_{0.02}$)(Si$_{1.88}$Al$_{0.12})$O$_6$
&Fiskenaesset, Greenland\\ &Bramble
Enstatite&(Mg$_{1.7}$Fe$_{0.28}$Ca$_{0.01}$)Si$_2$O$_6$&Bramble,
Norway\\ &&&\\ Clinopyroxene&Diopside&
(Ca$_{0.992}$Na$_{0.028}$Mg$_{0.94}$Fe$^{2+}_{0.017}$Fe$^{3+}_{0.013}$Cr$_{0.002}$Al$_{0.015}$)Si$_2$O$_6$&Dekalb,
NY, USA\\ &&&Smithsonian sample R18682\\
\end{tabular}
\end{minipage}
\end{table*}
Bulk samples of the minerals listed in Table \ref{tab:samples} were
broken into smaller pieces with a hammer and then ground to powders by
hand with a ceramic pestle and mortar. Grain size was measured by
visual inspection in a binocular microscope at $\times 63$
magnification. Small quantities of powder were mixed with petroleum
jelly on 0.8-mm-thick polyethylene substrates. Preliminary room
temperature measurements at Washington University indicated that the
grain sizes and the numbers of grains on each substrate were
sufficient to resolve most infrared bands. At Queen Mary, University
of London (QMUL) the samples and substrates were cut down and placed
in brass holders. The 295~K and 3.5~K measurements for each mineral
were performed with a single sample and the same spectrometer. The
only difference between the measurements was the positioning of the
sample in the instrument beam and the temperature of the
sample. Hence, the substrate, the coverage of the substrate and the
column density were \emph{identical} for each measurement of a
specimen.

\subsection{Cryogenics and Spectroscopy}
Spectra were obtained at QMUL using a Martin-Puplett polarising
Fourier Transform Spectrometer with a mercury arc lamp source and a
liquid-helium-cooled bolometer pumped to 1.5~K. For the room
temperature (295~K) measurements, the sample was placed in the output
beam of the spectrometer; the 3.5-K tests were performed by putting
the samples in a filter wheel in front of the detector within the
1.5-K bath in the cryostat. Since the sample was heated by incident
radiation from the arc lamp, it was slightly warmer than the bath, at
$3.5\pm0.5$K. The spectral resolution was selected by inspection of
the interferogram; to resolve the fine structure of end-member
forsterite at 3.5~K the resolution must be increased to
0.25~cm$^{-1}$, whilst a resolution of 2~cm$^{-1}$ is adequate at
295~K. For consistency, all spectra at all temperatures were obtained
at the higher resolution of 0.25~cm$^{-1}$.

For each temperature, reference spectra of the open aperture and
petroleum jelly on a polyethylene slide were obtained and then divided
into the spectrum of the sample to remove instrumental and substrate
effects. For the 3.5~K transmittance spectra, there is an uncertainty
in the flux calibration. Therefore the 3.5~K data were normalised to
the level of the 295~K transmittance spectra at 500~$\mu$m
(20~cm$^{-1}$) where the temperature dependence is small. The
transmittance scale of the room temperature spectra is better than
1$\%$ and the uncertainty in the normalisation of the 3.5~K data is
$\sim \pm10 \%$. The data were then converted to absorbance units
($A=-\ln{(Transmittance)}$).  Ripples in the spectra at wavelengths
$\ga 50~\mu$m are an artifact of imperfections in the thickness and
the orientation of the samples and references relative to the beam.

Data were obtained for thick films of forsterite and olivine and for
thinner films of olivine, and the pyroxenes. In the thick films, bands
shortwards of 45~$\mu$m were frequently saturated (these bands were
excluded from the figures and the analysis). The olivine and enstatite
thin films were too thin for bands longwards of 45~$\mu$m to be
resolved. The Bramble enstatite and diopside films were of sufficient
thickness for the majority of bands to be seen at both temperatures.
Since the mass and grain size distribution of the silicate powder is
unmeasured, the derived absorbances do not indicate mass density and
the absorbance per unit area per unit mass cannot be estimated from
these data at present.

\subsection{Astronomical Data}
For comparison of our laboratory data with SWS and LWS spectra from
ISO, we present observations of CPD -56$^\circ$8032, a planetary
nebula whose ISO spectrum was published by Cohen et al. (1999).  We
have reduced SWS observation 13602083 and LWS observation 08401538 and
subtracted a polynomial continuum fit (with no physical significance)
to show up the fine structure. Cohen et al. obtained a qualitative
match to their spectrum with room-temperature forsterite
(Mg$_2$SiO$_4$), orthopyroxene and clinopyroxene, using laboratory
data by Koike et al. (1993) multiplied by 65K (forsterite,
clinopyroxene) and 90K (orthopyroxene) blackbody continua. Features at
43$\mu$m and 62$\mu$m were matched with crystalline water ice. The
wavelengths and FWHMs of the silicate peaks in their fits are marked
in Figures~\ref{bowey_fig2} and \ref{longrel} by horizontal bars and
vertical ticks. (It is beyond the scope of this paper to deduce the
mineralogy of this source using our new laboratory data.)

\section{RESULTS}
\subsection{Overview of the laboratory dataset}
\begin{figure*}
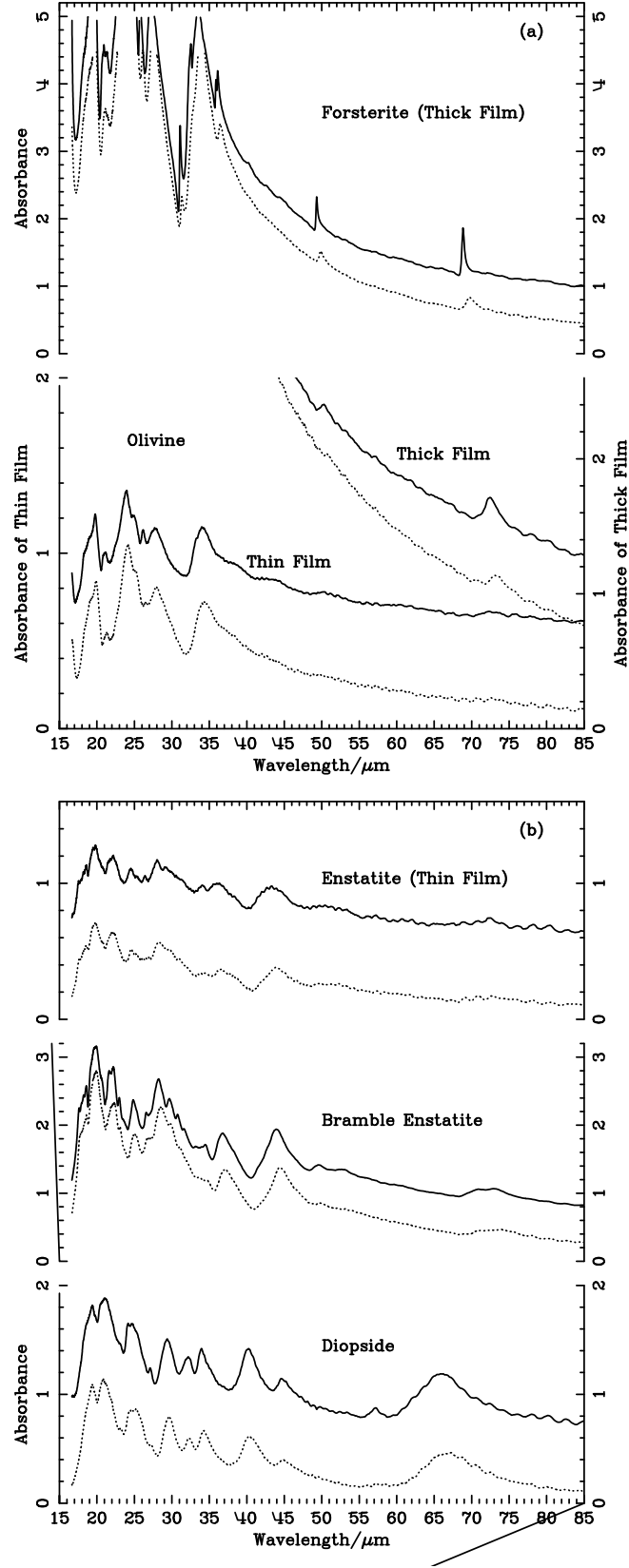

\epsfig{file=ma96603a.eps,width=0.49\linewidth,clip=,bbllx=115,bburx=520,bblly=228,bbury=740 }\\
\epsfig{file=ma96603b.eps,width=0.49\linewidth,clip=,bbllx=115,bburx=520,bblly=228,bbury=740 }
\caption{Overview of data: 15 to 85~$\mu$m absorbance spectra of (a)
olivines and (b) orthopyroxenes (enstatite and Bramble enstatite) and
a clinopyroxene (diopside) at 295~K (dotted) and 3.5~K (solid). Each
3.5~K spectrum is displaced by 0.5 units above the 295~K spectrum of
the same sample. For \emph{thick films}, bands shortwards of 45$~\mu$m
were saturated; for \emph{thin films}, bands longwards of 45$~\mu$m
were too weak to be resolved; the Bramble enstatite and diopside films
were of intermediate thickness so that most bands were resolved at
both temperatures.\label{bowey_fig1}}
\end{figure*}

The 15-85-$\mu$m absorbance spectra of the olivines (forsterite and
olivine), orthopyroxenes (enstatite and Bramble enstatite) and a
clinopyroxene (diopside) are presented in Figure
~\ref{bowey_fig1}. \label{bowey:results}
Since it may be possible to estimate the grain size distribution and
column density by electron microscopy at a later date, we present the
measured values of absorbance for these samples\footnote{This
normalisation differs from that presented previously: Bowey et
al. (2000) re-normalised pairs of 3.5~K and 295~K absorbance spectra
to the absorbance of the highest room temperature peak.}.

In general, (i) the room temperature spectra look like low-resolution
versions of their 3.5~K counterparts, even though the same spectral
resolution was used for all these measurements; (ii) the relative
absorbance of neighbouring peaks changes as the temperature is
reduced; (iii) the wavelength of a peak decreases with temperature and
the magnitude of the shift increases with the wavelength of the band.
The effect of temperature is most noticeable in the forsterite
spectra, with the bands becoming particularly narrow and strong: this
difference with degree of ionic substitution is real since this is the
only mineral devoid of Fe$^{2+}$ or other cations substituting for
Mg$^{2+}$ in the crystal structure.

\subsection{Effect of sample thickness}
Comparison of the spectra of the ``thick'' and ``thin'' olivine films
reveals that thick samples (i.e. larger amounts of powder embedded in
the petroleum jelly) are required to see the comparatively weak 50 and
70~$\mu$m bands and this has resulted in the saturation of the most
prominent features of forsterite and thick olivine at wavelengths
$<40~\mu$m. Hence, saturated features have been excluded from the
presented datasets. The room temperature spectra of the thin enstatite
and thin olivine films contain no distinguishable features longwards
of 45~$\mu$m. The Bramble enstatite and diopside films are of
intermediate thickness since bands are not saturated and the
far-infrared peaks are clearly distinguishable from the underlying
ripple.

\subsection{Band splitting, band enhancements and changes in bandwidth}
\begin{figure*}
    \epsfig{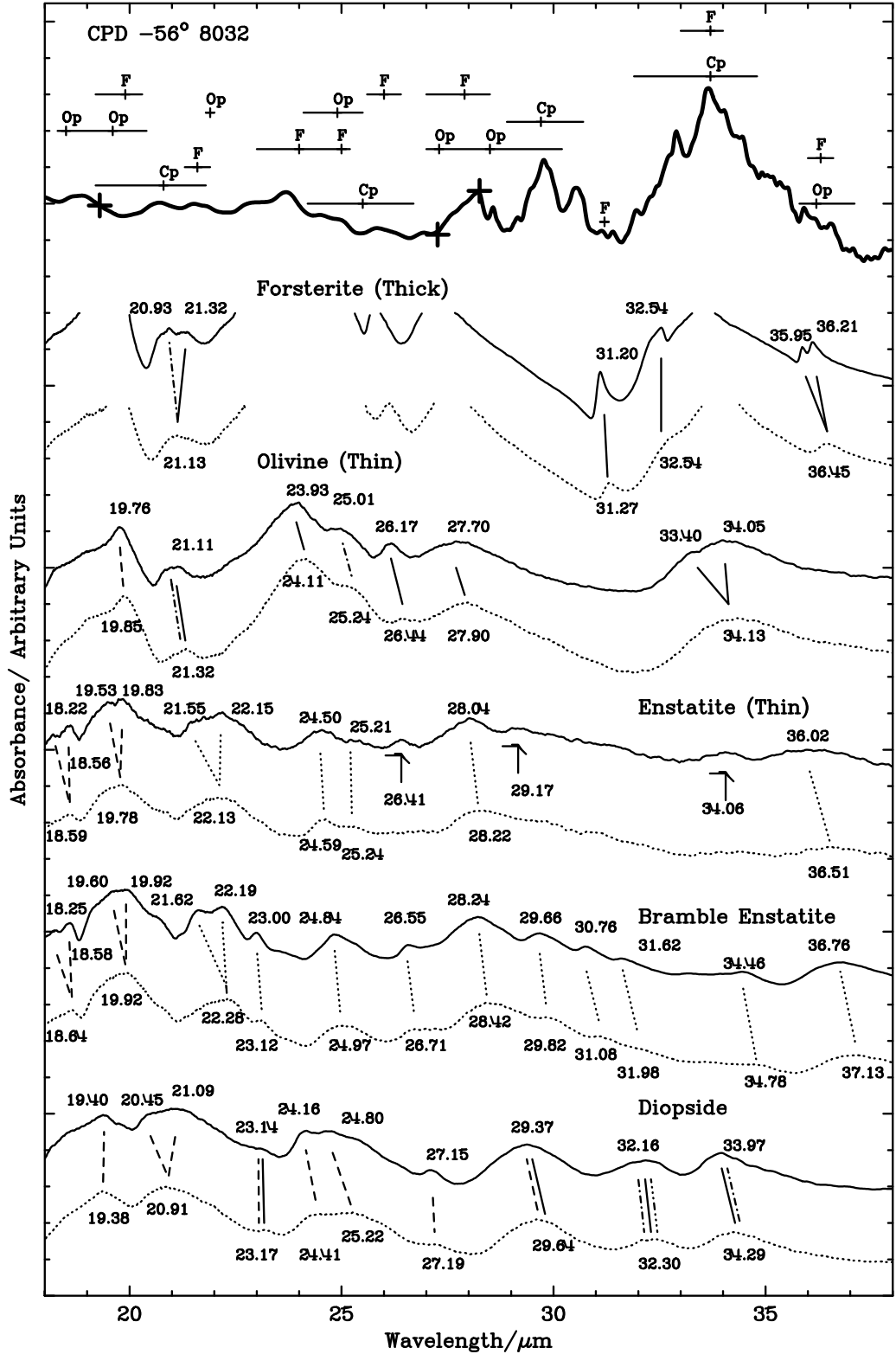}
\caption{Detailed structure in the 18--38-$\mu$m wavelength range for
the 295~K (\emph{dotted}) and 3.5~K (\emph{solid}) absorbance
measurements, with the ISO spectrum of CPD -56$^\circ$8032
(\emph{thick solid line}) plotted for comparison. {\bf ISO
observation:} \emph{bold crosses} indicate limits of spectral
segments; horizontal bars with ticks indicate the FWHM and
$\lambda_{pk}$ (respectively) of the features in the Cohen et
al. (1999) model obtained using room temperature data:
``F''=forsterite, ``Op''=orthopyroxene, ``Cp''=clinopyroxene). {\bf
Laboratory spectra:} Lines and pairs of lines linking pairs of spectra
respectively denote band shifts and band splitting.  The styles of
these lines indicate band assignments: \emph{solid}-- translation of
Mg$^{2+}$ and/or Fe$^{2+}$ cations, \emph{dashed}-- deformation of
Si--O--Si bands, \emph{dot-dash}-rotation of SiO$_4^{4-}$ tetrahedra,
\emph{dash-dot-dot-dot}-- translation of the silicate chain,
\emph{dotted}-- the assignments of these bands are unknown; arrows
denote bands seen only at 3.5~K. Offsets in the y-axis are arbitrary;
pairs of spectra are normalised to the same level; hence the relative
changes in band strength are real for individual
samples.\label{bowey_fig2}}
\end{figure*}

\begin{figure*}
  \begin{center}
    \epsfig{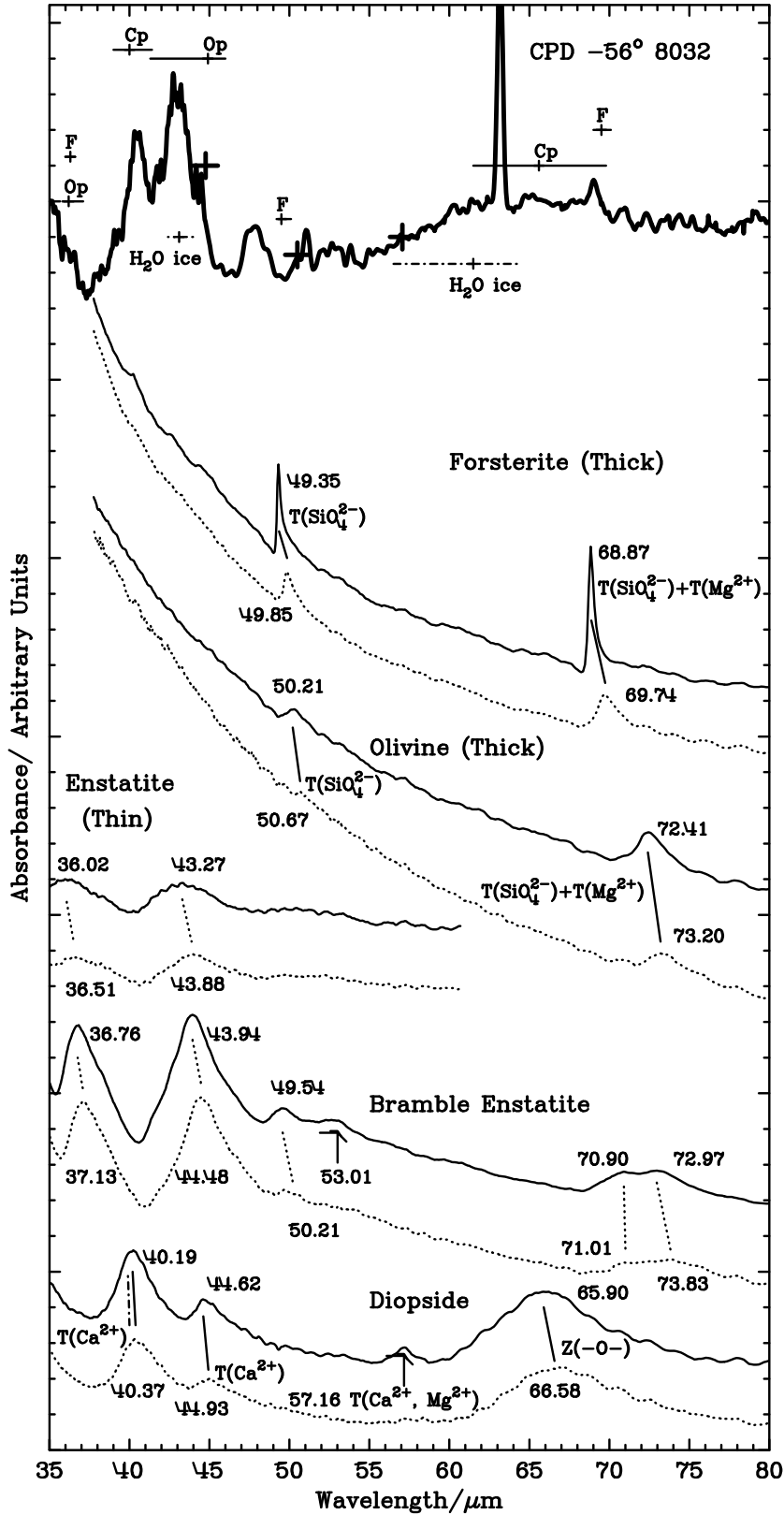}
    \end{center}
\caption{Band enhancements in the 38-80-$\mu$m wavelength range
compared with ISO data for CPD -56$^\circ$8032.  The \emph{solid}
lines linking the laboratory curves indicate band shifts due to ionic
translations, T(ion), except Z(--O--) which is a zig-zag motion of the
bridging oxygen atoms between silicate tetrahedra; \emph{dot-dash}
lines indicate rotations of SiO$_4^{2-}$ tetrahedra, band assignments
for \emph{dotted} lines are unknown. Arrows indicate extra bands
observed at 3.5~K. See Figure \ref{bowey_fig2} for key to remaining
symbols. \label{longrel}}
\end{figure*}

The detailed structure in the 18--38-$\mu$m and 38--80-$\mu$m regions
are shown in Figures~\ref{bowey_fig2} and \ref{longrel}, respectively.
Band splittings are seen shortwards of 38~$\mu$m; whilst wavelength
shifts, band enhancements and changes in the full-width-half-maxima
(FWHM) of peaks are seen throughout the 18-80-$\mu$m range.

\subsubsection{Band splitting}
In both olivines the degenerate $\sim 21~\mu$m Mg$^{2+}$ translation
and SiO$_4^{2-}$ rotation splits into two bands when cooled to 3.5~K
(Figure \ref{bowey_fig2}). The $\sim 34$-$\mu$m peak in forsterite is
saturated, however this Mg$^{2+}$ translation band is split in the
3.5-K olivine spectrum and the 36.45-$\mu$m 295~K Mg$^{2+}$
translation peak in forsterite is also split at 3.5~K. The band
splittings in forsterite are consistent with the peaks produced from
different polarisations occurring at the same frequency
\cite{jebhofmeister_1997}; at 295~K this is an unresolved doublet. In
the orthopyroxenes the band splitting is seen only in the three bands
shortward of 23~$\mu$m. Since the band assignments are undetermined in
the enstatites, the cause of the splitting is unknown. In the diopside
clinopyroxene only the degenerate Si--O--Si deformation at 20.91~$\mu$m
is split.
\subsubsection{Band enhancements}
The bands which are most strongly enhanced at low temperature, but
which occur in both the room temperature and the 3.5~K spectra, are
the unsaturated forsterite peaks, the 25.44-(295~K) to
26.17-$\mu$m~(3.5~K) olivine band, the 22.12-(295~K) to
23.00-$\mu$m~(3.5~K) band of Bramble enstatite and the
25.22-(295~K) to 24.80-$\mu$m~(3.5~K) band of diopside.

Some peaks which do not result from obvious splitting (or changes in
the substrate) are seen only in the 3.5-K spectra. Such features occur
in the 3.5-K thin enstatite spectrum at 26.41, 29.17 and
34.06~$\mu$m. Since similar bands are seen in Bramble enstatite
spectra obtained at both temperatures, we suggest that these features
are enhanced at 3.5~K. Other peaks are seen in Bramble enstatite but
not in enstatite (e.g. the 295-K 22.12-, 31.98-, and 31.08-$\mu$m
bands); this is probably due to the greater thickness of the Bramble
sample and to differences in the metal content in the two
minerals. Extra bands are seen in the 3.5~K spectrum of Bramble
enstatite at 53.91~$\mu$m, and in diopside at 57.16~$\mu$m.

\subsubsection{Changes in bandwidth}
Estimates of the full-width-half-maxima of peaks which are
well-separated from adjacent peaks and well-resolved are given in
Table \ref{tab:FWHM}; peaks which are unresolved doublets at 295~K or
are blended with stronger peaks have not been measured. The widths of
isolated singlet peaks provide insight into the population of the
vibrational modes. FWHM values were extracted with the {\sevensize IDL
GAUSSFIT} routine which approximates the continuum local to the peak
with a quadratic equation and fits a gaussian to the peak. As
described in Section \ref{sec:bandocc}, singlet peaks in the 3.5~K
spectra are narrower than those measured at 295~K. With the exception
of the forsterite peaks, bandwidths in the 3.5~K spectra are $\sim
90\%$ of their 295~K-widths, irrespective of the wavelength of the
peak. The 31-, 49- and 69-$\mu$m bands of 3.5~K forsterite are,
respectively, 90\%, 45\% and 31\% of their 295~K widths; these bands
arise from phonons along one direction within the lattice.
\begin{table}
\caption{Peak wavelengths ($\lambda_{pk}$ and FWHMs ($\Delta \lambda$)
of selected bands.}
\label{tab:FWHM}
\begin{tabular}{lcccc}
Sample&$\lambda_{pk}$~(295~K)&$\Delta\lambda$~(295~K)&$\lambda_{pk}$~(3.5~K)&$\Delta\lambda$~(3.5~K)\\
&$\mu$m&$\mu$m&$\mu$m&$\mu$m\\
Forsterite
&$31.27\pm0.02$&$0.29\pm0.02$ &$31.20\pm0.02$	&$0.26\pm0.02$\\
&$49.85\pm0.06$&$0.56\pm0.08$	&$49.35\pm0.06$&$0.25\pm0.08$\\
&$69.74\pm0.12$&$1.3\ \ \pm0.1\ $&$68.87\pm0.12$&$0.4\ \pm0.1\ \ $\\
Olivine
&$50.67\pm0.06$&-&$50.21\pm0.06$&$1.4\ \ \pm0.1$ \ \\
&$73.20\pm0.13$&$1.8\ \ \pm0.1\ $&$72.41\pm0.13$ &$1.6\ \ \pm0.1$\ \ \\
Bramble &$44.48\pm0.05$&$2.76\pm0.05$&$43.94\pm0.05$&$2.49\pm0.05$\\
Enstatite&&&&\\
Diopside
&$40.37\pm0.04$&$2.59\pm0.05$&$40.19\pm0.04$&$2.24\pm0.05$\\
&$44.93\pm0.05$&$1.77\pm0.05$&$44.62\pm0.05$&$1.63\pm0.05$\\
&$66.58\pm0.11$&$8.1\ \pm0.2\ \ $&$65.90\pm0.11$&$6.9\ \ \pm0.2$\ \ \\
\end{tabular}
The uncertainties in $\Delta\lambda$ include the effects of spectral
resolution, fringing and base-line estimates.
\end{table}
\subsection{Wavelength Shift and Splitting Patterns}
\begin{figure}
  \begin{center}
    \epsfig{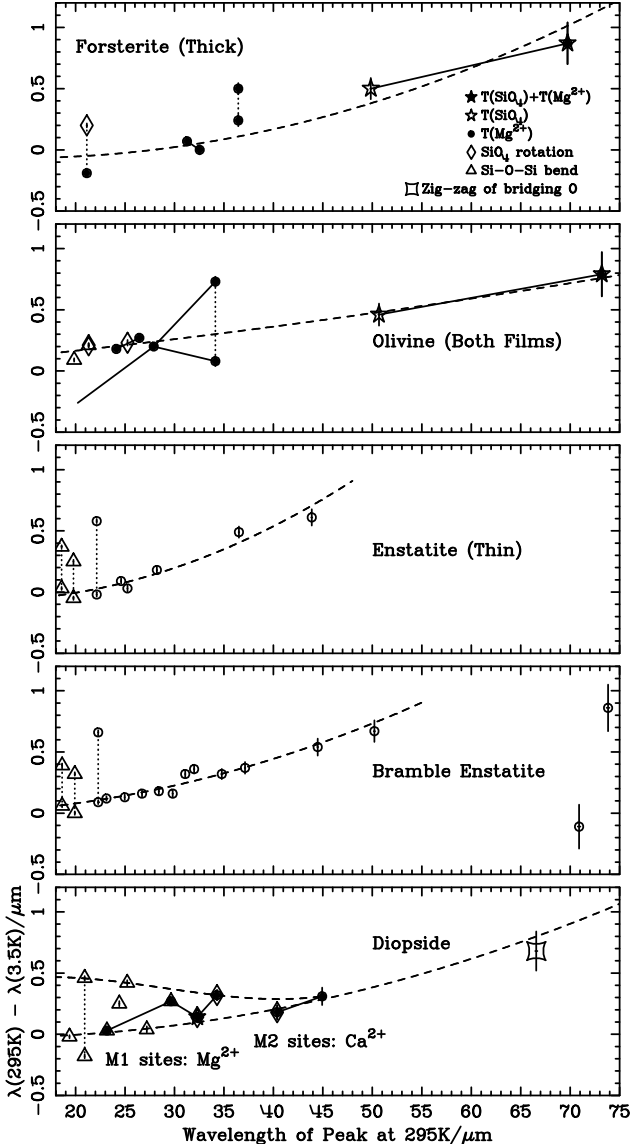}
\end{center}
\caption{Shifts in the wavelengths of the peaks ($\lambda(295$K$)-\lambda(3.5$K$)$) versus the wavelength of the peaks at 295~K ($\lambda(295$K$)$). Error bars are plotted where they are larger than the plotting symbols; empty \emph{circles} indicate unknown band assignments; \emph{dotted} lines indicate degenerate 295~K bands which split at 3.5~K; \emph{solid} lines connect bands which are produced by similar lattice components; \emph{dashed} lines indicate fits obtained in Section \ref{sec:shiftfit}. Some bands are missing from the forsterite and enstatite spectra because the samples were either too thick, or too thin for them to be resolved. \label{bowey_fig3}}
\end{figure}

In Figure \ref{bowey_fig3}, the shifts in the peak wavelength of the
bands between 295~K and 3.5~K, $\lambda(295~$K$)-\lambda(3.5~$K$)$,
are plotted as a function of the wavelength of the respective 295~K
peaks.
In general, bands shift to shorter wavelengths at 3.5~K and the
magnitude of the band shift increases with increasing wavelength, from
$\sim 0-0.2~\mu$m at 25~$\mu$m to $\sim 0.9~\mu$m at
80~$\mu$m. However, the shift patterns are quite complicated because
different peaks have different origins in the lattice. The situation
is complicated in some of these spectra since we have not been able to
resolve all the peaks in the thick forsterite and thin
enstatite. Measurements of the spectra of other film thicknesses are
required to determine the low-temperature behaviour of all the
infrared peaks.

The shift patterns and band splittings are similar within a solid
solution series if the lattice structure is the same. For example, the
$\sim 50$-$\mu$m and $\sim 70$-$\mu$m bands exhibited by forsterite
and olivine shift by similar amounts, --0.5~$\mu$m and --0.9~$\mu$m,
respectively, when the minerals are cooled from 295~K to 3.5~K. The
increase in the energies of these SiO$_4^{4-}$ and combined
SiO$_4^{4-}$ and Mg$^{2+}$ translations at 3.5~K indicates a
contraction of the lattice as the temperature is reduced; sharpening
of the bands is due to a reduction in phonon scattering at 3.5~K.

The ortho- and clinopyroxenes exhibit different shift and splitting
patterns. The band shifts in Bramble enstatite and in enstatite
increase fairly uniformly with wavelength. However shifts between
neighbouring bands in diopside clinopyroxene vary by up to 0.3~$\mu$m
and between 27 and 41~$\mu$m the pattern is almost saw-tooth. The
complex shift pattern of diopside is probably caused by the difference
in the size and mass of the Ca$^{2+}$ and Mg$^{2+}$ cations which
respectively occupy the M2 and M1 sites. Band splitting in the three
18--22.5-$\mu$m bands is common to both orthoenstatites-- the
Si--O--Si shifts in both minerals are $\sim 0.0$ and --0.4~$\mu$m and
the Mg$^{2+}$ shifts of the $\sim 22.5$-$\mu$m peak are 0.0 and
--0.5~$\mu$m. The splittings are caused by a reduction in the widths
of the peaks, and by different orientations (polarizations) of the
bonds within the lattice because the different crystallographic axes
contract with temperature at different rates.

\section{DISCUSSION}
\subsection{Astrophysical Implications}
These determinations of the relationship between temperature and band
shifts and band splitting could help in reaching a better
understanding of the mineralogy of astronomical regions showing
optically thin silicate bands in the far infrared. Such
temperature-dependent laboratory measurements could greatly benefit
the elucidation of the physical properties of dusty environments.

\subsubsection{Limits on the effect of temperature on $\lambda_{pk}$}
\label{sec:shiftfit}
The far-infrared bands observed in many astronomical environments are
wavelength-shifted with respect to bands of similar shape observed in
room-temperature laboratory spectra (Molster 2000). Wavelength shifts
can be introduced by temperature differences, but they can can also be
introduced by other factors such as chemical and isotopic composition
and degree of crystallinity. To place limits on the contribution of
temperature to the wavelength shift of a peak in astronomical data,
fits to the ratio,
$R=(\lambda(295$K$)-\lambda(3.5$K$))/\lambda(295$K$)$, for peaks which
did not split at 3.5K were obtained over the wavelength ranges
specified in Table \ref{tab:shiftfit}. Linear fits to $R$ were
obtained using the {\sevensize IDL LINFIT} routine; the fit to the
upper limit of the diopside shifts was obtained with the {\sevensize
IDL POLYFITW} routine. Polynomial fits to the wavelength shifts of
singlet bands between 295K and 3.5K are plotted in Figure
\ref{bowey_fig3} and the coefficients of these fits are are listed in
Table \ref{tab:shiftfit}.
\begin{table}
\begin{minipage}{\linewidth}
\caption{Fits to $R x=\lambda(295$K$)-\lambda(3.5$K$)$ using
polynomial fits of the form, $R=a+bx+cx^2$, where
x=$\lambda_{pk}(295$~K$)$, for the wavelength shift data plotted in Figure
\ref{bowey_fig3}.  }
\label{tab:shiftfit}
\begin{tabular}{lcccc}
Mineral&Range&$a$&$b$&$c$\\ &$\mu$m &$\times 10^{-3}$&$\times
10^{-3}$&$\times10^{-3}$\\ Forsterite&31-70&-9.66&0.347&-\\
Olivine&24-74&7.47&0.0396&-\\ Enstatite&24-44&-14.0&0.686&-\\ Bramble
Enstatite&23-52&-3.13&0.356&-\\ Diopside $L$\footnote{Lower boundary
of wavelength shifts: fit to 295~K peaks at 19.38, 23.17, 27.19,
32.30~$\mu$m and all peaks where
$\lambda>40~\mu$m.}&19-67&-5.44&0.262&-\\ \qquad\qquad
$U$\footnote{Upper boundary of wavelength shifts: fit to 295~K peaks
at 25.22, 34.29, 44.93~$\mu$m}&25-45&62.4&-2.55&0.0294\\
\end{tabular}
\end{minipage}
\end{table}

\subsubsection{Determination of grain temperatures}
\begin{figure} \begin{center}
\epsfig{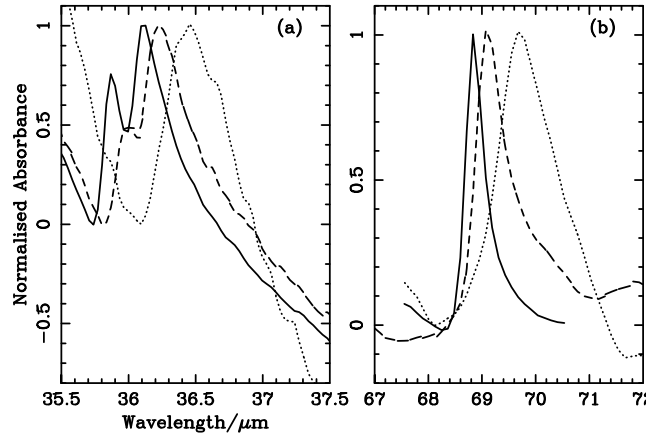}
\end{center}
\caption{Comparison between 3.5-K (\emph{solid}), $77\pm
5$-K(\emph{dashed}) and 295-K (\emph{dotted}) forsterite
spectra). 77~K spectra were obtained by placing the forsterite sample
in a cold finger cooled by liquid nitrogen within the output beam of
the spectrometer. \label{fig:69miceg}}
\end{figure}

Our results indicate that suitable high resolution spectra obtained at
a range of cryogenic temperatures can be used to determine the
characteristic temperatures of specific grain populations in
astronomical spectra. An emission feature which peaks near 69~$\mu$m
in circumstellar spectra has been identified with forsterite (Malfait
et al. 1998). However, the astronomical feature is shifted relative to
room temperature laboratory data, which give $\lambda_{pk}=69.74~\mu$m
(Figure~\ref{longrel} and Table~\ref{tab:FWHM}). In Figure
\ref{fig:69miceg}(b) we compare the shape of the 69-$\mu$m band of our
laboratory forsterite spectra obtained at three different
temperatures.  The spectral resolution of the ISO LWS grating
spectrometer at this wavelength was 0.3$\mu$m, compared to the
laboratory feature's FWHM of 1.3~$\mu$m at 295~K and 0.4~$\mu$m at
3.5~K, while the LWS wavelength precision was 0.03$\mu$m, compared to
the shift of 0.9$\mu$m in the peak wavelength between 295~K and
3.5~K. Therefore the derivation of characteristic grain temperatures
from ISO spectra is quite feasible (see also Molster 2000) and will be
reported on in a subsequent paper (Bowey et al., in preparation).
Band splitting and band shifts at other wavelengths, for example the
splitting in the 36.5$\mu$m forsterite band shown in Figure
\ref{fig:69miceg}(a), could also be used to determine grain
temperatures if the spectral resolution and signal-to-noise of the
astronomical data were sufficiently high. The relation between the
derived characteristic grain temperatures at different wavelengths may
aid the determination of the spatial distribution of the dust.

\subsubsection{Effect on inferred mineralogical compositions and dust masses}
The absorption coefficients of the forsterite bands are enhanced at
3.5~K compared to room temperature values. In contrast, there is
little change in the absorption coefficients of most of the pyroxene
features. Therefore, the true mass ratio of forsterite to enstatite
could be smaller than estimated using room-temperature laboratory
data. More mass is required to reproduce the strengths of many
features in ISO spectra using room temperature optical constants, than
would be required at the likely temperatures of the astronomical
grains. Measurements of the temperature-dependence of the mass
absorption coefficients are required in order to obtain the true
relative mineralogical composition and dust masses.

\section{SUMMARY}
We have obtained 0.25~cm$^{-1}$ resolution 15--85~$\mu$m spectra of
powdered minerals embedded in petroleum jelly at $295\pm2$~K and
$3.5\pm0.5$~K.  The samples were two olivines, forsterite
(Mg$_2$SiO$_4$) and olivine var. peridot
Mg$_{1.77}$Fe$_{0.20}$Al$_{0.01}$Ni$_{0.01}$SiO$_4$, two
orthoenstatites, enstatite and Bramble enstatite, and a clinopyroxene
(MgCa diopside).\\

In general:

\begin{enumerate}
\item the room temperature spectra look like lower-resolution versions
of their 3.5~K counterparts. In both of the olivines the degenerate
$\sim 21 ~\mu$m Mg$^{2+}$ translation and SiO$_4^{2-}$ rotation splits
into two bands when cooled to 3.5~K. The $\sim 34$-$\mu$m peak in
forsterite is saturated, however this Mg$^{2+}$ translation band is
split in the 3.5~K olivine spectrum and the 36.45-$\mu$m 295~K
Mg$^{2+}$ translation peak in forsterite is also split at 3.5~K. In
the orthoenstatites, band splitting is seen only in the three bands
shortward of 23~$\mu$m. In the MgCa diopside clinopyroxene only the
20.91-$\mu$m degenerate Si--O--Si deformation is split. Different
sample thicknesses must be measured to verify that no splitting is
seen in other bands.\\

\item the wavelength of a peak decreases with temperature and the
magnitude of the shift increases with the wavelength of the band, from
$\sim 0-0.2\mu$m at 25~$\mu$m to $\sim 0.9~\mu$m at
80~$\mu$m. However, the shift patterns are quite complicated and
depend on lattice structure as well as stoichiometry since they differ
between ortho- and clino-pyroxenes. For example, the band shifts in
Bramble enstatite and enstatite increase with wavelength. However, in
the diopside clinopyroxene the shifts between neighbouring bands vary
by up to 0.3~$\mu$m -- between 27 and 41~$\mu$m the pattern is almost
saw-tooth.  The situation is complicated in these spectra since we
have not been able to resolve all the peaks at all temperatures. For
the olivines and enstatites the wavelength shifts of bands which are
singlets at 295K and 3.5K can be approximated by polynomials of the
form $ax+bx^2$ where $x=\lambda_{pk}(295$~K$)$ and coefficients $a$
and $b$ differ between minerals. Lower limits to the wavelength shifts
for diopside can be obtained using polynomials of this form, however a
third order term is required to obtain the upper estimates for
wavelength shifts in the 20--41$\mu$m region.\\

\item bands sharpen as the temperature is reduced.  At 3.5~K the
widths of isolated single bands in the olivine, enstatites and
diopside are $\sim 90\%$ of their 295~K-widths. However, in
forsterite, the 3.5-K, 31, 49 and 69-$\mu$m bandwidths are,
respectively, 90\%, 45\% and 31\% of their 295~K widths.\\

\item the relative absorbances of neighbouring peaks change and bands
sharpen as the temperature is reduced; this effect is most noticeable
in the forsterite spectra. Peaks at 53.91~$\mu$m in Bramble
enstatite and at 57.16~$\mu$m in diopside appear only in the 3.5~K
spectra.
\end{enumerate}

Our results show that comparison between the band shifts, bandwidths
and band splitting in laboratory data obtained at a range of
temperatures with astronomical far-infrared spectra can allow a
determination of the characteristic temperatures of crystalline grains
responsible for the optically thin emission features. Further
low-temperature spectra at a range of temperatures are required to
ascertain the precise temperature dependence of the shift, since the
relationship between $T$ and $\lambda_{pk}$ is unlikely to be linear
between 295K and 3.5K and it is possible that the bands stop shifting
at some temperature above 3.5K. Ultimately it may be possible to
determine the temperature of distinct grain populations by comparison
of the observed peak wavelengths and FWHMs with laboratory spectra.
Since the absorption coefficients of bands in laboratory spectra also
vary with temperature, new measurements may allow better estimations
of grain masses and silicate mineralogy. Additional high-resolution
laboratory spectra of a larger variety silicates, at a larger number
of temperatures, are desirable for comparison with ISO data and with
future measurements obtained by spectrometers on board SIRTF, FIRST
and the NGST. These laboratory measurements are planned.

\section*{acknowledgements}
CL is supported by a Royal Society Research Fellowship; JEB and CT are
supported by PPARC; AMH is supported by NSF under grant
AST-9805924. The Bramble enstatite sample was supplied by R.F. Dymek
(Washington U.) and the diopside sample was supplied by the
Smithsonian Museum. STARLINK, IDL and ISAP software were used for data
reduction and analysis. We thank the anonymous referee for a
constructive report which has improved this paper.

\label{lastpage}
\end{document}